\documentclass[twocolumn, showpacs,preprintnumbers,amsmath,amssymb, superscriptaddress, prl]{revtex4}

\usepackage{graphicx}
\usepackage{dcolumn}
\usepackage{bm}

\begin{document}

\title{Odd- and even-order dispersion cancellation in
quantum interferometry}

\author{Olga Minaeva}
\affiliation{
Dept. of Electrical \& Computer Engineering, Boston University, Boston, Massachusetts 02215}

\affiliation{
Department of Physics, Moscow State Pedagogical University, 119992 Moscow (Russia)}

\author{Cristian Bonato}
\affiliation{
Dept. of Electrical \& Computer Engineering, Boston University, Boston, Massachusetts 02215}

\affiliation{
CNR-INFM LUXOR, Department of Information Engineering, University of Padova, Padova (Italy)}

\author{Bahaa E.A. Saleh}
\affiliation{
Dept. of Electrical \& Computer Engineering, Boston University, Boston, Massachusetts 02215}

\author{David S. Simon}
\affiliation{
Dept. of Electrical \& Computer Engineering, Boston University, Boston, Massachusetts 02215}

\author{Alexander V. Sergienko}
\affiliation{
Dept. of Electrical \& Computer Engineering, Boston University, Boston, Massachusetts 02215}

\affiliation{
Dept. of Physics, Boston University, Boston, Massachusetts 02215}

\begin{abstract}
We describe a novel effect involving odd-order dispersion
cancellation. We demonstrate that odd- and even-order dispersion
cancellation may be obtained in different regions
of a single quantum interferogram using frequency-anticorrelated
entangled photons and a new type of quantum interferometer. This
offers new opportunities for quantum communication and metrology
in dispersive media.
\end{abstract}

\pacs{03.67.Bg, 42.50.St, 42.50.Dv, 42.30.Kq}

\maketitle

\section{INTRODUCTION}

The even-order dispersion cancellation effect based on
nonclassical frequency-anticorrelated entangled photons has been
known in quantum optics for some time \cite{franson92,
steinberg92a}. The nonlinear optical process of spontaneous
parametric down conversion (SPDC) traditionally provides a
reliable source of frequency-entangled photon pairs with
anticorrelated spectral components, as a consequence of energy
conservation.  If the frequency of the signal photon is
$\omega_s$, then the frequency of its twin idler photon must be
$\omega_i=\Omega_p - \omega_s$, where $\Omega_p$ is the frequency
of the pump beam. A quantum interferometer records the modulation
in the rate of coincidence between pulses from two photon-counting
detectors at the output ports of a beamsplitter in response to a
temporal delay between two spectrally correlated photons entering
its input ports symmetrically. This type of quantum optics
intensity correlation measurement, exhibited in the Hong-Ou-Mandel
(HOM) interferometer \cite{hom87},
 is manifested by an observed dip in the rate of coincidences.
In previous demonstrations of dispersion cancellation, one photon
of the downconverted pair travels through a dispersive material in
one arm of the HOM interferometer while its twin travels only
through air. The final coincidence interference dip is not
broadened in this case, demonstrating insensitivity to even-order
dispersion coefficients \cite{steinberg92a,qoct02}.

Even-order dispersion cancellation has been used in  quantum
information processing, quantum communication, and in quantum
optical metrology. For example, it enhances the precision of
measuring photon tunneling time through a potential barrier
\cite{steinberg92b} and improves the accuracy of remote clock
synchronization \cite{giovannettiNAT01}. The same effect provides
superior resolution in quantum optical coherence tomography
\cite{nasr03} by eliminating the broadening of interference
envelope resulting from group velocity dispersion.
The potential of quantum even-order dispersion cancellation has recently stimulated efforts to mimic this effect by use of classical nonlinear optical analogues \cite{shapiroOCT06,resch07, reschNature08}.

In this Letter we introduce a novel type of quantum interferometer that enables demonstration of the odd-order dispersion cancellation as a part of new dispersion management technique. In our design,  both even-order and odd-order dispersion cancellation effects can be recorded as parts of a single quantum interference pattern.

\begin{figure} [ht]
\centering
\includegraphics [width = 8 cm] {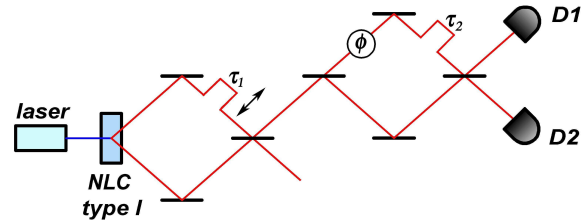}
\caption{Schematic diagram of the optical setup. The SPDC source
produces pairs of frequency anticorrelated photons combining on a
beamsplitter in a HOM interferometer configuration. Photons
exiting one HOM port are fed into a MZ interferometer. Coincidence
events are registered between two single-photon detectors at the
output ports of the MZ interferometer. A dispersive sample in one
arm of the MZ interferometer generates a phase delay ($\phi $).}
\label {setup}
\end{figure}

HOM interferometers are commonly used to produce either $ \left |
\Psi \right  \rangle \sim   \left | 2,0 \right  \rangle - \left |
0,2  \right  \rangle $ state,  when the delay $\tau_1$ is set to
balance the two paths, ensuring destructive interference in the
middle of the interference dip, or a superposition of $ \left |
1,1 \right  \rangle$, $ \left | 0,2  \right  \rangle $ and $ \left
| 2,0  \right  \rangle $ states, when the delay $\tau_1$
significantly unbalances two paths and shifts coincidences to the
shoulder of HOM interference pattern. Mach-Zehnder (MZ)
interferometers fed by a particular quantum state have also been
studied in detail \cite{Campos90}.

In the new design two interferometers work together: one output
port of a HOM interferometer provides input to a MZ
interferometer. The state of light introduced into the MZ
interferometer is continuously modified when the delay $\tau_1$ in
the HOM interferometer is scanned. A signal from one of the HOM
output ports is fed into a MZ interferometer with a dispersive
sample providing a phase shift $ \phi$ in one arm, as shown in
Fig.1. The delay $\tau_2$ inside the MZ interferometer is kept at
a fixed value. A peculiar quantum  interference pattern is
observed in the rate of coincidences between two photon-counting
detectors $D1$ and $D2$ at the output ports of the MZ
interferometer as a function of $\tau_1$. The interference profile
has two distinct patterns.  The central interference pattern
depends only on even-order dispersion coefficients, while the
peripheral pattern depends only on odd-order terms. This ability
to manipulate and evaluate odd-order and even-order dispersion
terms independently in a single quantum interferometer opens new
perspectives in quantum communication and in precise optical
measurement.

\section{THEORETICAL MODEL}

For detectors $D_1$ and $D_2$ much slower than the temporal
coherence of the downconverted photons, the coincidence rate
in such intensity correlation measurements is \cite{SPDC_theory}:

\begin{equation}
\label{eq:Rc}
R_{c}(\tau_{1}, \tau_{2}) = \int dt_{1}\int dt_{2} G^{(2)}(t_{1},t_{2}),
\end{equation}
\noindent with $G^{(2)}(t_1,t_2)$ second order correlation
function $G^{(2)}(t_1,t_2)$:
\begin{equation}
\label{eq:G2}
G^{(2)}(t_{1},t_{2}) = \mid \left \langle 0 \right | \hat{E}_{1}^{(+)}(t_{1}) \hat{E}_{2}^{(+)}(t_{2}) \left | \Psi \right \rangle \mid^{2}.
\end{equation}
\noindent $ E_{1}^{(+)}(t_{1})$ and $ E_{2}^{(+)}(t_{2})$ are the
electrical field operators at the surfaces of detectors $D_1$ and
$D_2$, respectively.
\begin{equation}
\label{eq:fields}
\hat{E}_{j}^{(+)}(t_{j)} = \frac{1}{\sqrt{2\pi}}\int d\omega_{j}e^{-i\omega_{j}t_{j}}\hat{b}_{j}(\omega_{j}),
\end{equation}
\noindent where $\hat{b}_{j}(\omega_{j})$ is the mode operator at
detector $j$, expressed in terms of the input field operators
$\hat{a}_{j}(\omega_{j})$  \cite{SPDC_theory}. The quantum state
of light emitted in a frequency-degenerate non-collinear type-I
phase-matching SPDC process with a monochromatic pump $\Omega_{p}$
is:
\begin{equation}
\label{eq:state}
\left | \Psi \right \rangle \propto\int d\omega f(\omega) \hat{a}_{1}^{\dagger}(\Omega_{0}+\omega)\hat{a}_{2}^{\dagger}(\Omega_{0}-\omega) \left | 0 \right \rangle,
\end{equation}
\noindent where $f(\omega)$ is a photon wavepacket spectral
function defined by the phase matching condition in the nonlinear
material,  $\Omega_{0} = \Omega_{p}/2$ is a central frequency of
each wavepacket, $ \omega_{s}= \Omega_{0}+\omega$ is the signal
photon frequency, and $ \omega_{i}=\Omega_{0}-\omega$ is the idler
frequency .

The phase shift $\phi (\omega)$ acquired by the broadband optical
wavepacket as it travels through a  dispersive material could be
expanded in a Taylor's series \cite{book_Diels_Rudolf_ultrashort}:

\begin{equation}
\label{eq:dispersion}
\phi(\omega_{s,i}) = c_0 + c_1 (\omega_{s,i}-\Omega_{0}) + c_2( \omega_{s,i}-\Omega_{0})^2+c_3( \omega_{s,i}-\Omega_{0})^3 + \cdots
\end{equation}

\noindent where the linear term $c_1$ represents the group delay
 and the second-order term $c_2$ is responsible  for group delay
dispersion. In a conventional white-light interferometer, $c_1$ is
responsible for a temporal shift of the interference pattern
envelope, $c_2$ causes its temporal broadening, while $c_3$
provides a non-symmetric deformation of the wavepacket envelope.
Higher-order terms might be included when a strongly dispersive
material is used  or in the case of extremely broadband  optical
wavepackets.

In the optical setup of Fig.1, the dispersive material providing
phase shift $\phi(\omega)$ could be situated in three possible
locations. When the sample is placed an arm of the HOM
interferometer it leads to the well-known even-order dispersion
cancellation effect \cite{qoct02}.  It may be shown that the
presence of a dispersive material between the two interferometers
does not affect the coincidence interferogram. We thus concentrate
on the most interesting case: we place the dispersive sample of
phase shift $\phi(\omega)$ inside the MZ interferometer, with
delay $\tau_2$ set to a fixed value, and $\tau_1$ as the variable
parameter.

Following the usual formalism \cite{SPDC_theory}, one can show
that the coincidence rate between the detectors is:
\begin{equation}
\label{eq:final}
\begin {split}
R_{c}(\tau_{1}, \tau_{2}) =& \int d\omega (\Phi_{0} - \Phi_\alpha (\omega, \tau_{2}) - \Phi_\beta (\omega,\tau_{2}))\cdot \\
&(f(\omega)f^{*}(\omega) + f(\omega)f^{*}(-\omega)e^{-2 i \omega \tau_{1}} ) ,
\end {split}
\end{equation}
\noindent where  $ \Phi_{0}$ is a constant,
\begin{equation}
\label{eq:odd}
\begin {split}
\Phi_\alpha (\omega,\tau_{2})= &e^{-2 i \omega \tau_{2}} e^{i \phi  (\Omega_{0} - \omega)} e^{-i \phi  (\Omega_{0} + \omega)} + c.c.,
\end {split}
\end{equation}
\noindent and
\begin{equation}
\label{eq:even}
\begin {split}
\Phi_\beta (\omega,\tau_{2})= &e^{-2 i \Omega_{0} \tau_{2}} e^{-i \phi  (\Omega_{0} - \omega)} e^{-i \phi  (\Omega_{0} + \omega)} + c.c.
\end {split}
\end{equation}

Although not obvious from the form of equation (\ref{eq:final}),
$R_{c}(\tau_{1}, \tau_{2})$ is a real function for any spectrum
$f(\omega)$, as can be seen by rewriting Eq. (\ref{eq:final}) in
manifestly real form:

\begin{eqnarray}
\label{eq:final2} R_{c}(\tau_{1}, \tau_{2})&=&\int d\omega\left\{
|f(\omega )|^2 +|f(-\omega )|^2 \right. \nonumber
\\&& \qquad +\left.\left[ e^{-2i\omega \tau_1}f(\omega )f^\ast (-\omega )+c.c.\right]\right\} \nonumber\\&& \qquad \qquad\times \left[ \Phi_0 -\Phi_\alpha (\omega
)-\Phi_\beta (\omega )\right]
\end{eqnarray}
This fact ensures that the technique demonstrated here applies to
all types of broadband frequency-anticorrelated states of light,
including those with nonsymmetric spectral profiles produced in
chirped periodically-polled nonlinear crystals.

The final coincidence counting rate $R_{c} (\tau_{1},\tau_{2})$ of
Eq. (\ref{eq:final}) may also be written as a linear
superposition:

 \begin{equation}
\label{eq:coincidence2}
R_{c} (\tau_{1},\tau_{2}) = B+ R_{0}(\tau_{1})-R_{even}(\tau_{1},\tau_{2})-R_{odd}(\tau_{1},\tau_{2}).
\end{equation}
\noindent The first coefficient B incorporates all terms that are
not dependent on the variable delay $\tau_{1}$, providing a
constant after integration. It establishes a baseline level for
the quantum interfererogram. The following terms:
\begin{equation}
\label{eq:term1} R_{0}(\tau_{1}) = 4 \int d\omega
f(\omega)f^{*}(-\omega)e^{-2 i \omega \tau_{1}},
\end{equation}
\begin{equation}
\begin {split}
R_{even}(\tau_{1},&\tau_{2})=\int d\omega f(\omega)f^{*}(-\omega)\cdot\\
e^{-2 i \omega \tau_{1}}& [ e^{-2i\Omega_{0} \tau_{2}} e^{-i \phi  (\Omega_{0} - \omega)} e^{-i \phi  (\Omega_{0} + \omega)} \\
&+ e^{2 i \Omega_{0} \tau_{2}} e^{i \phi  (\Omega_{0} - \omega)}
e^{i \phi  (\Omega_{0} + \omega)}],
\end {split}
\end{equation}
\begin{equation}
\begin {split}
R_{odd}(\tau_{1},&\tau_{2})= \int d\omega f(\omega)f^{*}(-\omega)\cdot\\
&[ e^{-2 i \omega (\tau_{1}+\tau_{2})} e^{i \phi  (\Omega_{0} - \omega)} e^{-i \phi  (\Omega_{0} + \omega)} +\\
&e^{-2 i \omega (\tau_{1}-\tau_{2})} e^{-i \phi  (\Omega_{0} - \omega)} e^{i \phi  (\Omega_{0} + \omega)}]
\end {split}
\end{equation}
\noindent are responsible for the shape of the interference pattern.

The term $R_0 (\tau_1)$ represents a peak centered at $\tau_1=0$
that is simply a Fourier transform of the down converted radiation
spectrum and is insensitive to the dispersion associated with
$\phi(\omega)$. Since $R_{even} (\tau_1,\tau_2)$ is dependent on
the sum $\phi  (\Omega_{0} - \omega) + \phi  (\Omega_{0} +
\omega)$, it is sensitive only to even-order terms in the
expansion Eq. (\ref{eq:dispersion}). This manifests odd-order
dispersion cancellation and generates a dispersion-broadened
function centered around $\tau_1 = 0$.  The last term $R_{odd}
(\tau_1,\tau_2)$,  in contrast, is sensitive only to odd-order
dispersion terms in $\phi(\omega)$. This term demonstrates the
well known even-order cancellation. The coefficients  $e^{-2 i
\omega (\tau_{1}+\tau_{2})}$ and $e^{-2 i \omega
(\tau_{1}-\tau_{2})}$ shift the two dips away from the center of
the interference pattern in opposite directions. Such
decomposition  of quantum interference terms makes it possible to
observe odd-order  and even-order dispersion cancellation effects
 in two distinct regions of the coincidence
interferogram.

\section{EXAMPLE}

Our results are illustrated by a numerical example of quantum
interference for a 3-mm thick slab of a strongly-dispersive
optical material ZnSe, inserted in one arm of the MZ
interferometer to provide the phase shift $\phi(\omega)$.  In this
experiment we assume the use of frequency-entangled down-converted
photons with 100-nm wide spectrum. As illustrated in Fig. 2, one
can identify the narrow peak $R_0 (\tau_1)$ in the center, which
is insensitive to dispersion, along with the component $R_{even}
(\tau_1,\tau_2)$, which is broadened by even-order dispersion
contributions only.  This central component of the interferogram
illustrates the odd-order dispersion cancellation effect.

\begin{figure} [ht]
\centering
\includegraphics [width = 8 cm] {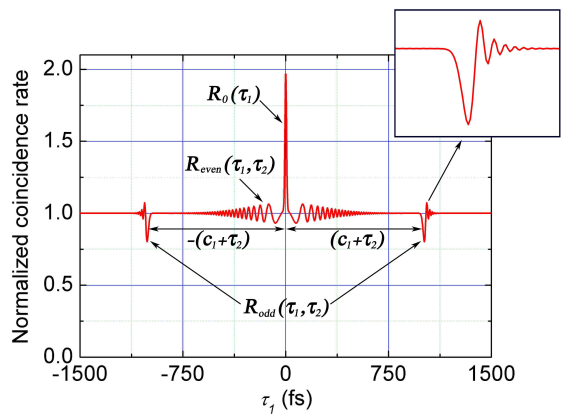}
\caption{The normalized coincidence rate as a function of
$\tau_{1}$ when a 3-mm thick ZnSe sample is placed in the MZ
interferometer. The fixed delay  $\tau_{2} = 26 $ ps is used. The
insert illustrates the odd-order dispersion contribution.} \label
{ZnSe}
\end{figure}

Two symmetric side dips  $R_{odd} (\tau_1,\tau_2)$ appear shifted far
away from the central peak by the group velocity delay
$c_1$ acquired by entangled photons
inside the dispersive material. However, this shift can be controlled by properly adjusting
the value of the fixed delay $\tau_{2}$.  Such a simple adjustment
moves both dips back closer to the center and makes it convenient
for  observing both dispersion cancellation features in a single scan of  the variable delay line $(\tau_1)$ inside HOM interferometer (see Fig. 2). The appearance of
asymmetric fringes on the side of two dips is a clear sign of the
third-order dispersion.
\cite{book_Diels_Rudolf_ultrashort}.

\section{DISCUSSION}

This result can also be understood physically by analyzing all
possible probability amplitudes  that lead to measured coincidence
events between $D1$ and $D2$.  The MZ interferometer input is a
pair of spectrally-entangled photons separated by time delay
$\tau_1$; if the leading photon has a high frequency, the lagging
photon will have a low frequency, and vice-versa.  We consider
first the case when no dispersive element is present, so that the
MZ interferometer introduces only a time delay $\tau_2$ between
its two arms. We assume that $\tau_2$ is much greater than the
photon wave packet width, $\tau_c$.  To explain the dependance of
the photon coincidence rate on $\tau_1$, as shown in
Fig.\ref{ZnSe}, we consider three processes occurring at the input
ports of the last beam splitter in the MZ interferometer:

1) If $|\tau_1|> \tau_c$  and  $|\tau_2 - \tau_1| >\tau_c$, then
the two photons arriving at the final beam splitter will be
distinguishable, so that no quantum interference is exhibited.

2) If $ |\tau_1| \approx |\tau_2|$, so that $|\tau_2-\tau_1| <
\tau_c$, then quantum interference can occur when the leading
photon takes the long path of the MZ interferometer and the
lagging photon takes the short path. The two arrive almost
simultaneously (within a time $ \tau_c$) at the two ports of the
final beam splitter.  Then the Hong-Ou-Mandel (HOM) effect is
exhibited at the beam splitter, albeit with only 25\% visibility
because of the presence of the other possibility that both photons
arrive at a single port, leading to a background coincidence rate
independent of  $\tau_1$. From a different perspective, one may
regard this scenario as similar to that obtained in a Franson
interferometer \cite{Franson89}, for which photon pairs follow
long-long or short-short paths. This scenario explains the
components of the coincidence interferogram near $\tau_1 = \pm
\tau_2$, and in this case the two spectrally-entangled photons
entering separate ports of the final beam splitter lead to quantum
interference accompanied by even-order dispersion cancellation.

3) Finally, when $|\tau_1| < \tau_c$, then one possibility is that the photons
arrive at separate input ports of the final beam splitter.  Since these photons are
separated by a time $\tau_2 >> \tau_c$, they are distinguishable and do not
contribute to quantum interference.  The other possibility is that the pair arrive
at the same beam splitter input port. In this case, upon transmission or reflection
at the beam splitter there are two alternatives for producing coincidence: transmission of the
high-frequency photon and reflection of the low-frequency photon, or vice-versa. This explains
the component of the coincidence interferogram near $\tau_1 \approx 0$.
In this scenario, which involves two spectrally-entangled photons entering a single
port of a beam splitter, quantum interference is accompanied by odd-order dispersion cancellation.
We thus see that the quantum interference effects
exhibited in scenarios 2) and 3) are accompanied by dispersion cancellation --
although in opposite manners in the two cases.

In conclusion, we have demonstrated a new effect in which even-
and odd-order dispersion cancellations appear in different regions
of a single interferogram. This is achieved via
frequency-anticorrelated photons in a new quantum interferometer
formed by a variable delay HOM interferometer followed by a
single-input, fixed-delay Mach-Zehnder interferometer. The
possibility of independently evaluating even- and odd-order
dispersion coefficients of a medium has potential for applications
in quantum communication and in quantum metrology of complex
dispersive photonics structures. In particular, the ability to
accurately characterize higher-order dispersion coefficients is of
great interest in the study of flattened-dispersion optical fibers
\cite{ferrando2001,reeves2002} and in dispersion engineering with
metamaterials \cite{elef2005}. The demonstrated potential of
even-order dispersion cancellation has stimulated the search for
classical analogues \cite{shapiroOCT06,resch07}. We expect that
the scheme presented here would also trigger the similar
development of nonlinear optical techniques mimicking this quantum
effect. Finally, note that our apparatus may be extended by adding
a second Mach-Zehnder to the unused HOM output port, allowing the
investigation of new four-photon interference effects.

\section{ACKNOWLEDGMENTS}
We would like to thank Andrey Antipov from SUNY Buffalo for assistance with numerical simulations. This work was supported by a U. S.
Army Research Office (ARO) Multidisciplinary University Research Initiative (MURI) Grant; by the Bernard M. Gordon Center for Subsurface
Sensing and Imaging Systems (CenSSIS), an NSF Engineering Research Center; by the Intelligence Advanced Research Projects Activity (IARPA)
and ARO through Grant No. W911NF-07-1-0629.

\end{document}